\documentclass[
    ,final            
  ,numberedheadings 
  ]
  {aipproc}

\layoutstyle{8x11double}
\newcommand{\tal}{\it et al. \rm}
\newcommand{\AAA}{{A\&A}} 

\newcommand{\ApJ}{{ApJ}}

\newcommand{\AJ}{{AJ}}

\newcommand{\MN}{{MNRAS}}
\newcommand{\MNRAS}{{MNRAS}}

\begin{document}

\title{Formation and dynamical evolution of galaxies and of  their components}


\author{E. Athanassoula}{
  address={Observatoire de Marseille, 2 Place Le Verrier, 13248
  Marseille C\'edex 4, France}
}
\begin{abstract}
From this vast subject, I will pick out and review three specific
topics, namely the formation and evolution of bars, the formation
of bulges, and the evolution during multiple major mergers.

Bars form naturally in galactic discs. Their evolution is driven by
the exchange of angular momentum within the galaxy. This is emitted
mainly by near-resonant material in the inner disc (bar), and is
absorbed by near-resonant material in the outer disc and in the halo. 
As a result of this, the bar becomes stronger and rotates slower.

Bulges are not a homogeneous class of objects. Based on their
formation history, one can distinguish three types. Classical bulges
are mainly formed before the actual disc component, from collapses
or mergers and the corresponding dissipative processes. Boxy/peanut
bulges are parts of bars seen edge-on. Finally, disc-like bulges
are formed by the inflow of material to the center due to bar torques.

Major mergers bring strong and fast evolution and can turn discs 
into ellipticals. I present results from simulations of multiple
mergers in groups of disc or of elliptical galaxies and discuss the
orbital anisotropy in the merger remnant. 
\end{abstract}

\maketitle


\section{Introduction}

Since reviewing all the formation and evolution processes of 
galaxies and of their components is an impossible task within the      
limits set by the time and page allocation of an invited review,
I will concentrate here on 
three specific topics, namely the formation and evolution
of bars, the formation of bulges and the evolution during multiple 
major mergers.

The formation and evolution processes can be distinguished into
fast and slow processes, and into internally and externally driven
processes (see also Kormendy \& Kennicutt 2004). Fast processes occur
on time-scales 
comparable to dynamical times, while slow processes, often referred
to as secular processes, have much longer time-scales, of the order
of a few Gyrs. Of course, there is no sharp distinction between the
two. Major mergers and collapses are prime examples of fast processes,
and have been associated with the formation of elliptical galaxies
and of some types of bulges. On the other hand, the evolution of bars
is a much slower process, with time-scales of the order of Gyrs.

Formation and evolutionary processes can also be distinguished, by their
origin, into internally and externally driven processes. Mergers,
both minor and major, as well as accretion, are externally driven.
On the other hand, collapses, or the secular evolution of bars are
internally driven. In some cases, like the formation of spirals
and bars, both internal and external processes can be invoked.
Obviously, in groups and clusters the effects of the environment are even
stronger than for field galaxies.

\section{Bar formation and evolution}
\label{sec:bars}

Bars are near-ubiquitous in disc galaxies, as shown by recent
near-infrared observations (Eskridge \tal 2000). Furthermore, the
fraction of bars with moderate to high strength seems to be
near-constant over the last 8 Gyrs (Sheth \tal 2003,
Jogee \tal 2004, Elmegreen, Elmegreen \& Hirst 2004). 

\begin{figure}
  \includegraphics[height=.5\textheight]{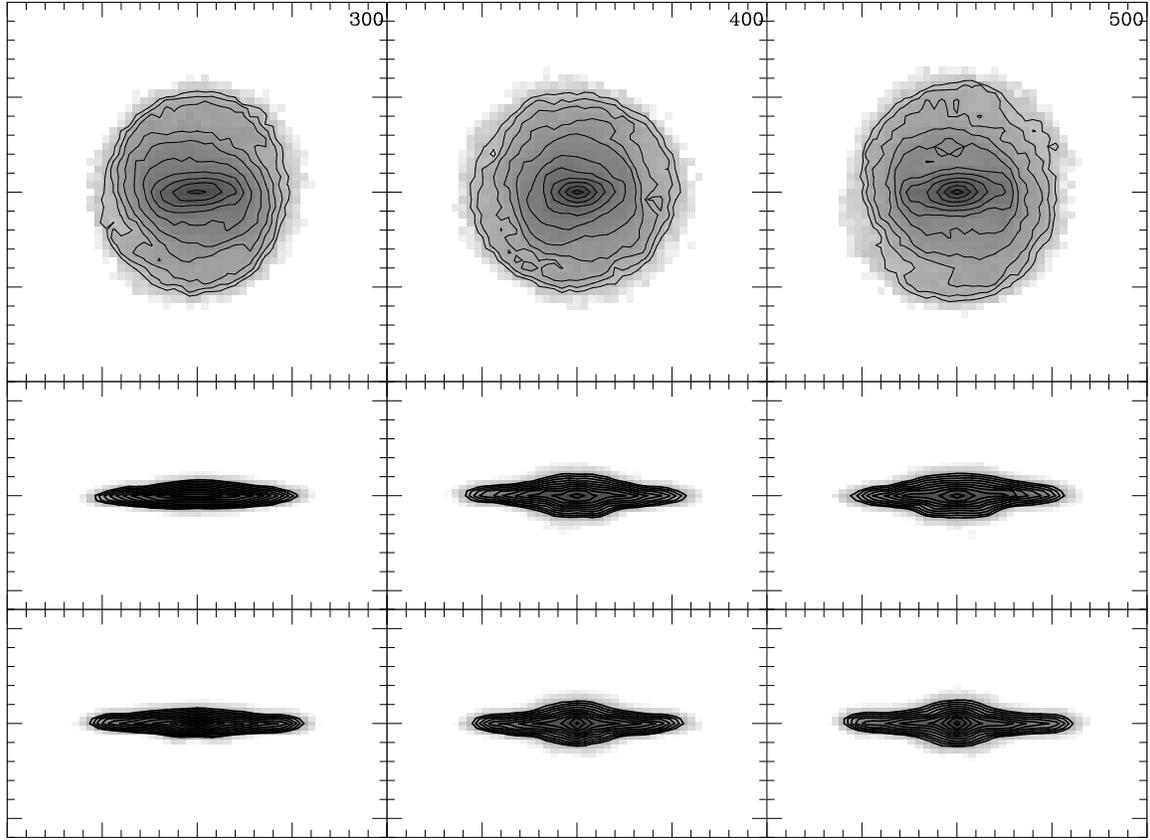}
  \caption{Formation of a bar in an initially axisymmetric disc. 
  The upper row gives 
  the face-on views; the second one the side-on views (i.e. edge-on
  with the line of sight along the bar minor axis) and
  the third row the end-on views (i.e. edge-on
  with the line of sight along the bar major axis). Time
  increases from left to right and is given in
  the upper right corner of each face-on panel. Here and elsewhere in 
  this section, I use computer units as defined by 
  Athanassoula \& Misiriotis (2002).}
\label{fig:evolA}
\end{figure}

\begin{figure}
  \includegraphics[height=.5\textheight]{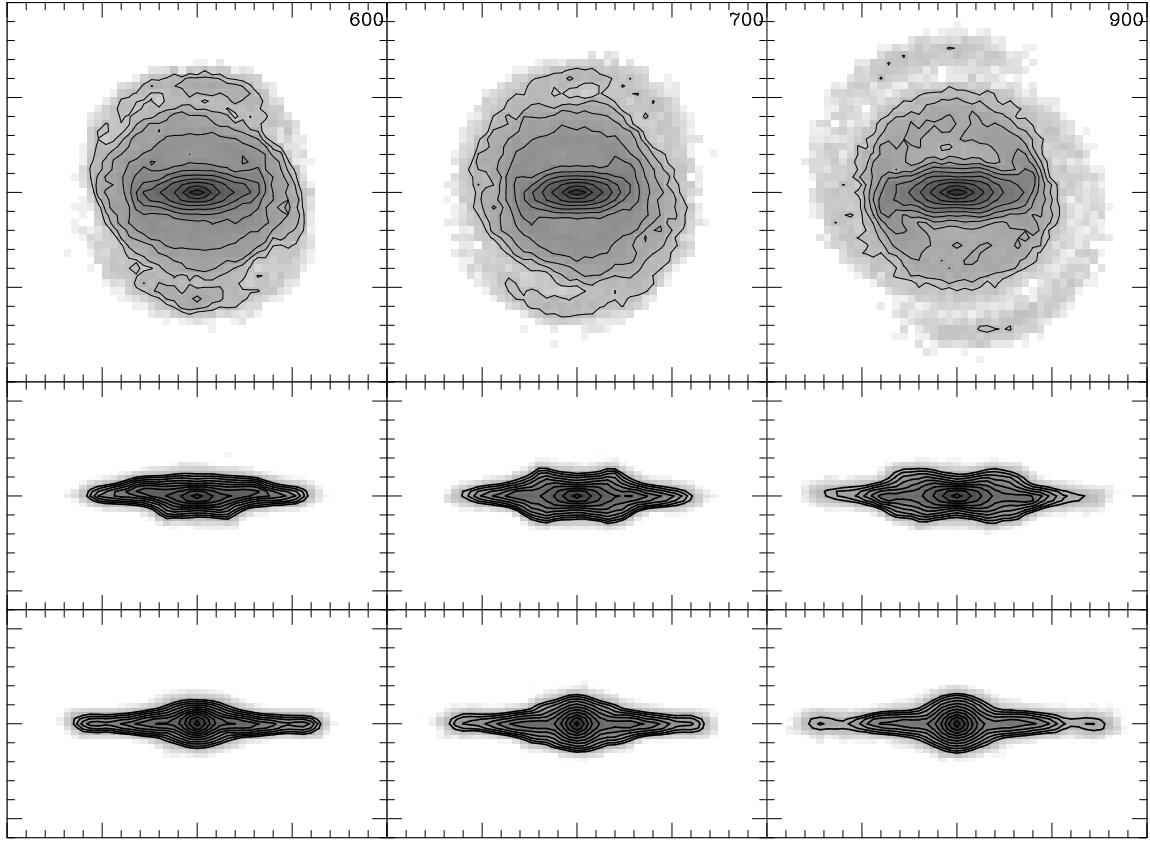}
  \caption{As in Fig.~\ref{fig:evolA}, but for later times. Note that
  both the disc extent and the bar strength increase considerably with
  time.}
\label{fig:evolB}
\end{figure}

An example of the formation and evolution of a bar component is given
in Figures~\ref{fig:evolA} and \ref{fig:evolB}. It is obtained from a
$N$-body simulation, similar to those described in Athanassoula and
Misiriotis (2002) and Athanassoula (2003). The disc, starting from 
axisymmetric at $t$ = 0 (not shown), forms a bar relatively early on
in the simulation. This bar evolves with time and becomes stronger
(mainly longer and more massive). Its isodensity contours
evolve from elliptical-like to rectangular-like and it acquires a ring
which surrounds it, reminiscent of the inner rings in barred
galaxies (Athanassoula \& Misiriotis 2002). Viewed side-on
(i.e. edge-on with the line of sight along 
the bar minor axis) the bar thickens considerably in the inner parts and by
the end of the simulation has formed a characteristic peanut
shape. Viewed end-on (i.e. edge-on with the line of sight along
the bar major axis) it also develops a central thick component,
of the same shape as classical bulges (see
subsection~\ref{subsec:classicalb} below for the definition). Other
simulations, from different initial conditions, show an evolution
which is qualitative similar, but which can differ considerably
quantitatively, i.e. the evolution can occur at different
time-scales and lead to bars of considerably different strength.

\begin{figure}
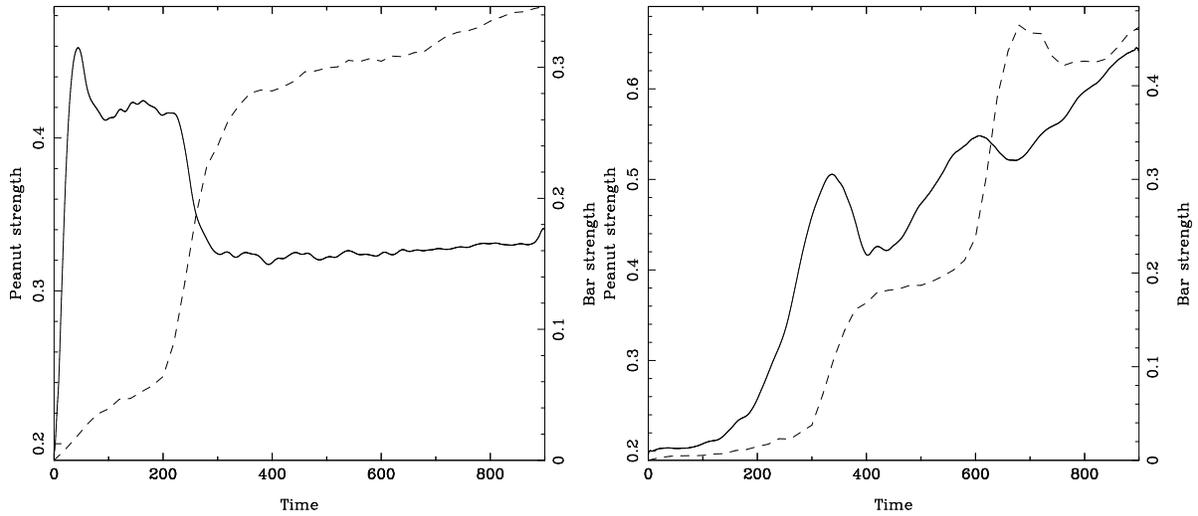

  \includegraphics[height=.3\textheight]{bar_peanut_nde037.ps}
  \includegraphics[height=.3\textheight]{bar_peanut_kdb008.ps}
  \caption{Evolution of the bar strength (solid line) and of the peanut
  strength (dashed line) with time for two simulations. In the left
  panel, note that the peanut strength
  increases abruptly with time roughly between times 200 and 300, and
  that the bar strength decreases during the same time interval. The
  simulation shown in the right panel has two such
  episodes. The first one is between times 300 and 400 and the second
  one between 
  times 600 and 700. }
\label{fig:strength}
\end{figure}

This evolution is due to the exchange of angular momentum within the
galaxy. This is emitted by material at near-resonance in the inner
disc (bar region) and particularly at the inner Lindblad resonance
(see Figure 1 of Athanassoula 2003). It is absorbed by near-resonant
material in the outer disc and particularly in the halo (Athanassoula
2002, 2003). As a result, the bar grows stronger (as witnessed in
Figures~\ref{fig:evolA} and \ref{fig:evolB}). 

A closer look at Figure~\ref{fig:evolA} shows that this bar growth is not
necessarily monotonic over the whole evolution time. E.g. the bar is
less strong at time 400\footnote{I use here the computer units
  introduced in Athanassoula \& Misiriotis (2002). By the calibration
of that same paper, the computer time unit is 1.4 10$^7$ yrs. Thus, time
400 corresponds to 5.6 10$^9$ yrs.} than at time 300, and then gets stronger
again. To follow this quantitatively, I compare in
Figure~\ref{fig:strength} the run of the bar strength and of the 
peanut strength with time, for two different simulations. In the first
case (left panel) the bar grows very abruptly before time
50\footnote{The maximum around time 50 is due to an episode of spiral
  formation.} and then stays nearly at constant amplitude till about
time 200. It then drops abruptly up to about time 300 and then shows
only a very mild increase with time. It is interesting to compare this
with the evolution of the peanut strength (same figure). This increases
very abruptly between times 200 and 300, i.e. at the same times when
the bar strength decreases. Such a drop of the bar strength can already
be seen in Figure 4 of Athanassoula (2002) and 
in Figures 2 and 4 of Martinez-Valpuesta, Shlosman \& Heller (2005), while  
the antagonism between bar and peanut strength was discussed by
Raha \tal (1991), Martinez-Valpuesta \& Shlosman (2004) and
Martinez-Valpuesta \tal (2005). 

This episode of coupled peanut formation and bar decrease need not be
unique in the bar evolution history, as discussed in Martinez-Valpuesta
\tal (2005) and as shown in the right panel of
Figure~\ref{fig:strength}. Here we see 
clearly two such episodes, one between times 300 and 400 and the
second between times 600 and 700. 

\section{Bulges}

Two different definitions have been used so far for
bulges. Accordingly, the bulge is either

\begin{enumerate}
\item
a smooth light distribution that
swells out of the central part of a disc viewed edge-on, or
\item
the extra light in the central part of the disc, above the
extrapolated exponential profile fitting the main part of the disc.
\end{enumerate}
Unfortunately, these two definitions are neither equivalent, nor even
consistent (see e.g. Bureau \tal 2005 for a discussion). Furthermore,
the thus classified `bulges' form a very inhomogeneous class of of
objects. Based on their formation scenarios, Athanassoula (2005)
distinguished three different types of bulges :

\subsection{Classical bulges}
\label{subsec:classicalb}

These are formed by gravitational collapse or by hierarchical merging
of smaller objects and the corresponding dissipative gas 
processes. The formation process is generally fast and sometimes
externally driven. It occurs early on in the galaxy formation process,
before the present disc was formed. Several versions of this
scenario have been elaborated (Carlberg 1984a,b; Steinmetz \& M\"uller
1995; Steinmetz \& Navarro 2002; Sommer-Larsen \tal 2003; Noguchi
1999; Immeli \tal 2004; Kauffmann 1996;
Kauffmann, Charlot \& White 1996). 
An alternative view is that classical bulges formed after the present disc,
by accretion, e.g. of a small elliptical (Pfenniger 1993; Athanassoula
1999; Aguerri, Balcells \& Peletier 2001; Fu, Huang \& Deng 2003).

Bulges formed in this way should have several similarities to elliptical 
galaxies, including their photometric radial profiles, their
kinematics and their stellar populations (e.g. Davies \tal 1983; Franx
1993; Wyse, Gilmore \& Franx 1997; and references therein). Thus, they
should 
be composed of predominantly old stars, they should have predominantly  
ellipsoidal shapes and should have near-$r^{1/4}$ projected density
profiles. A typical object in this category is e.g. the Sombrero 
galaxy (NGC 4594). 

\subsection{Box/peanut bulges}
These objects form during the
dynamical evolution of barred galaxies. As discussed in 
section~\ref{sec:bars}, bars form naturally and relatively fast in
disc galaxies and then evolve at a slower rate. The time necessary for
the initial bar formation is  
of order of a few galaxy rotations and, more precisely, depends on the
halo-to-disc mass ratio within the main body of the galaxy. As shown
in Figures~\ref{fig:evolA} and \ref{fig:evolB}, somewhat after bar
formation some of the material in the bar acquires stronger vertical
motions and thus reaches larger distances from the equatorial
plane. These distances increase with time. Viewed edge-on, this gives a
characteristic box/peanut shape. Thus, box/peanut bulges are simply
{\it parts} of bars seen edge-on. 

Objects formed in this way should have observed morphological,
photometrical and kinematical properties that are the same as those of
$N$-body bars seen edge-on (Kuijken \& Merrifield 1995; Merrifield \&
Kuijken 1999; Athanassoula \& Bureau 1999; Bureau \&
Freeman 1999; Chung \& Bureau 2004; Bureau \&
Athanassoula 1999, 2005; Athanassoula 2005). Since
they form by rearrangement of disc material,
they should be constituted of stellar populations that are similar
to those of the inner disc at radii comparable to those of the
box/peanut feature. Subsequent star formation in one, or both, of
these components can introduce some young stars. The age of the
bulk of the stars, however, can be considerably older than the age of the
boxy/peanut feature itself. The average size of these features
should be of the order of 1 to 3 disc scale-lengths, and can not
reach $D_{25}$. This formation scenario has been well worked out in a number of
papers describing relevant $N$-body simulations (e.g.  Combes \tal
1990; Raha \tal 1991; Athanassoula \& Misiriotis 2002; Athanassoula
2003, 2005; O'Neill \& Dubinski 2003; Martinez-Valpuesta \& Shlosman 2004; 
Martinez-Valpuesta \tal 2005). Orbital structure theory predicts 
that box/peanut bulges should be shorter than the corresponding bars,
and this is indeed confirmed both by simulations and by observations (see
Athanassoula 2005 for a complete discussion of this issue).

\subsection{Disc-like bulges}
Contrary to boxy/peanut bulges, the
formation scenario of disc-like bulges is not fully worked out, but the general
picture is as follows : It is well known that gas will
concentrate to the inner parts of the disc under the influence of the
gravitational torque of a bar, thus forming an inner disc extending
roughly up to the (linear) inner Lindblad resonance, or forming a ring at such
radii (e.g. Athanassoula 1992). The extent of
this region is of the order of a kpc. When this 
disc/ring becomes sufficiently massive it will form stars, which
should be observable as a young population in the central part of
discs. Kormendy \& Kennicutt (2004) estimate that the star formation
rate density in this region is 0.1 - 1 M$\odot$ yr$^{-1}$ kpc$^{-2}$, i.e. 1
to 3 orders of magnitude higher than the 
star formation rate average over the whole disc. This will lead
naturally to the formation of a sizeable central disc. Note that
disc-like bulges can also be formed in $N$-body simulations with no
gas, from inward motions of the disc material. These disc-like
bulges, however, will be less massive than
those formed by combined stellar and gaseous processes. 

Disc-like bulges have properties attributed normally
to disc systems and can contain substructures found normally in
discs, mainly spirals, rings, bright star-forming knots, dust lanes and 
even bars, as discussed e.g. in Kormendy (1993), Carollo,
Stiavelli \& Mack (1998) and Kormendy \& Kennicutt (2004). They can
contain a sizeable amount of gas, as well as stars  
younger than those formed with the two previous scenarios, as well as
a fraction of old stars. They qualify as 
bulges by the second of the above given definitions, but {\it not}
by the first one. Their radial photometric profiles are closer to
exponential than to 
$r^{1/4}$ (e.g. Courteau, de Jong \& Broeils 1996; 
Carollo, Stiavelli \& Mack 1998). They are primarily found in late
type disc galaxies 
(e.g. Andredakis \tal 1995; Carollo \& Stiavelli 1998) as expected
since gas processes  can enhance their formation. Good examples,
however, have also been found in S0 galaxies (Erwin \tal 2003).
Their host galaxies often harbour a large-scale bar.
Prominent strong central peaks have indeed been observed in
radial photometric profiles obtained from cuts along the major axis of
edge-on barred galaxies (see e.g. Figure 1, 2 and 
5 in L\"utticke, Dettmar \& Pohlen (2000) and Figure 1 in Bureau \tal
(2006)). These features are only seen 
in cuts which are on, or very near the equatorial plane, which argues for
the disc-like geometry of the light distribution that produces them. 

The three types of bulges are very different but are not mutually
exclusive. Face-on strongly barred galaxies often harbour what
looks like classical bulges and should thus contain both a classical
and a box/peanut bulge. Furthermore, the strong central peaks of
the radial density profile in edge-on boxy/peanut systems
(e.g. L\"utticke \tal 2000; Bureau \tal 2006), which
I have linked to disc-like bulges in the
previous paragraph, show that boxy/peanut bulges often co-exist with
disc-like ones.

\section{Multiple major mergers and haloes of ellipticals}

The original suggestion (Toomre 1977) that ellipticals form by
mergers of disc galaxies has been followed by a number of
investigations
based partly on $N$-body simulations and partly on observations
(see Barnes and Hernquist 1992, and references therein). It is now
well established that disc galaxies are embedded in massive and
extended haloes (for a review, see Bosma (1999) and references
therein). During the merger of two, or more, disc galaxies into
an elliptical, the halo components should also merge, and form the
dark halo component of the merger remnant, the mass of which
should be equal to the sum of the masses of the haloes of the
progenitors. Thus, ellipticals should have a similar ratio of
total dark to total baryonic mass as discs. This was
corroborated by a number of 
observations, e.g. by X-ray or by gravitational lensing. It thus came as a
big surprise when measurements 
from PNe of the projected velocity dispersion in the outskirts
of some elliptical galaxies gave very low values, thus arguing
for little or no dark matter in them (Mendez \tal 2001; Romanowsky 
\tal 2003). 

Using $N$-body simulations, Dekel \tal (2005) argued that the
trajectories of stars in the outer parts could be very elongated, 
and thus lead to radially anisotropic velocity dispersions, which
could explain the low observed values of the projected $\sigma$.
This very interesting proposal was discussed during this meeting
both by K. Kuijken and by G. Mamon, and I refer the reader to 
their contributions for more information on this subject.

How general and how robust is the Dekel \tal prediction of radial orbits?
To elucidate this further, 
I will extend here this discussion to the case of multiple 
mergers, to which the formation of elliptical galaxies may also
be due (Weil and Hernquist 1994, 1996, Athanassoula \& Vozikis 1999). I will
present preliminary results on the merging of 5 identical galaxies, either
5 disc galaxies, or 5 ellipticals. These simulations have between a
couple and a few million particles and include neither gas nor star
formation. According 
to Dekel \tal the stars which formed during the merger should have
only about 10 to 15\% lower $\sigma$ than the older stars. The
presence of the gas, however, could influence the whole evolution. Our
purely stellar models should be adequate for simulating 
mergings between ellipticals and/or between early type disc galaxies.
On the other hand, the Dekel \tal simulations considered Sb-Sc galaxies 
as progenitors. The present simulations are thus in
all respects very different from those of Dekel \tal and it is 
interesting to check whether they also produce merger remnants with orbital
structure biased towards radial orbits.  
A more general discussion of these simulations will be presented
elsewhere (Athanassoula 2006, in preparation).

\begin{figure}
  \includegraphics[height=.3\textheight]{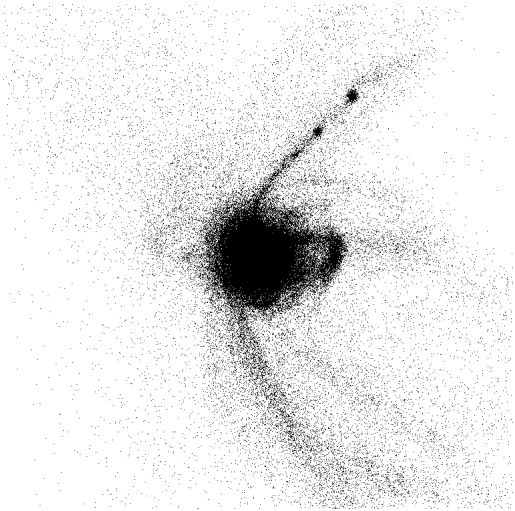}
  \caption{A multiple merging. Note the large number
of features, and in particular a rather spectacular tail-like feature
which contains a number of massive clumps which could later become
dwarf ellipticals.}
\label{fig:merger}
\end{figure}

Multiple mergers are dynamically more complicated than pair mergers.
Indeed, multiple mergers can be considered as a sequence of pair
mergers, of types S + S, E + E, or S + E. where S denotes a disc 
galaxy and E an elliptical. But they can also be near-simultaneous
mergers of all the initial galaxies, i.e. considered as an extreme
example of a clumpy collapse. Figure~\ref{fig:merger} shows the result 
shortly after a multiple merger and reveals the existence of many 
structures and particularly of many tails. Note the clumps formed 
in one of these tails, which could be the precursors of dwarf
ellipticals.

I measured the velocity dispersion radial profiles in such remnants
after their fast evolution has subsided and some quasi-equilibrium
has been reached. Calculating the same radial profiles at 
later times gives similar results (see also Fig. 8 of Dekel \tal
2005). I then calculated the anisotropy parameter $\beta$, defined
as 

\begin{equation}
\beta = 1 - 0.5 (\sigma_\theta^2 + \sigma_\phi^2) / \sigma_r^2,
\end{equation}

\noindent
where $\sigma_r$, $\sigma_\theta$ and $\sigma_\phi$ are the components
of the velocity dispersion in spherical coordinates.
Isotropic cases, with $\sigma_r$ = $\sigma_\theta$ =
$\sigma_\phi$ gives $\beta$ = 0. Models with purely circular
orbits ($\sigma_r$ = 0) have $\beta$ = -- $\infty$. Finally, models
with purely radial orbits ($\sigma_\theta$ =
$\sigma_\phi$ = 0) give $\beta$ = 1.

\begin{figure}
  \includegraphics[height=.3\textheight]{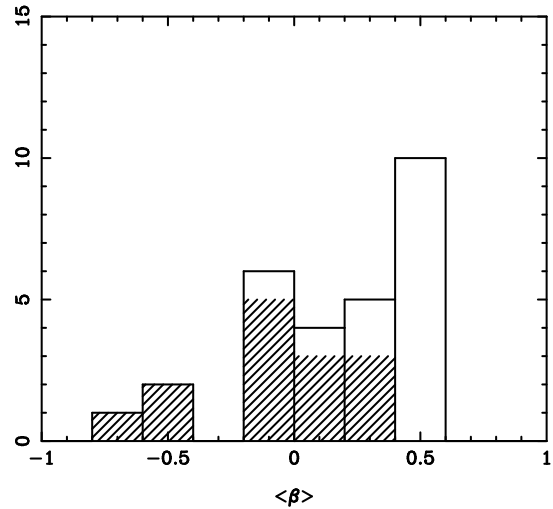}
  \caption{Histogram of the number of merger remnants with a given value of
$<\beta>$. The solid line refers to all the simulations and the 
hatched area to simulations starting with a group of five
identical ellipticals.}
\label{fig:histo}
\end{figure}

I then obtained for each simulation the average $\beta$ value,
$<\beta>$, for the region between 2 and 5 effective radii.
Figure~\ref{fig:histo} shows the histogram of these $<\beta>$ values.
Note that there is a spread of $<\beta>$ values between -- 0.8 and
0.6. This is much larger than the spread of values found by Dekel
\tal (2005) in their simulations of pair mergers (roughly between 0.2
and 0.75), and, particularly, contains examples with a large
fraction of near-circular orbits. To investigate this further I give,
in the same figure, the same histogram but now only the simulations
starting from a 
group of elliptical galaxies (hatched area in the figure). It is clear
that all but one of the simulations that give remnants with a velocity
distribution which is biased towards circular-like orbits, rather than
towards near-radial ones, have elliptical progenitors. On the other
hand, simulations starting with 
disc galaxies only behave in a way similar to that found by Dekel 
\tal for the merger remnants from two equal mass spirals; although the
$<\beta>$ values found here are, on average, somewhat
smaller. Note that Dekel 
\tal (2005) also discuss one simulation in which the progenitors had
initially a classical bulge and find that this reduces
$\beta$. The extrapolation of this result to mergers between 
ellipticals is in agreement with the results presented here.

\begin{figure}
  \includegraphics[height=.3\textheight]{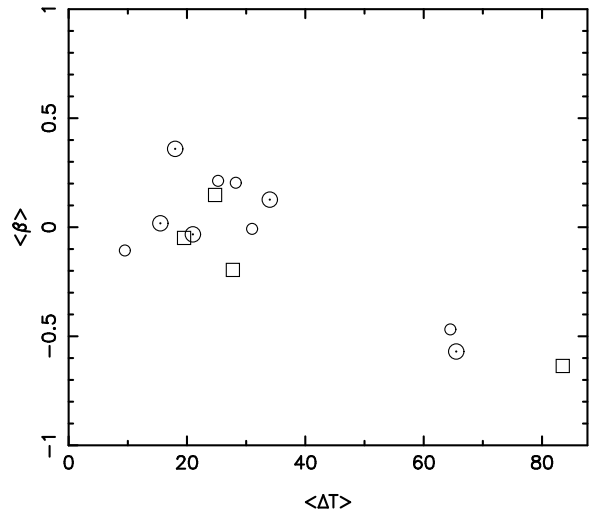}
  \caption{$<\beta>$ as a function of the mean time between successive
mergers. Each symbol denotes a simulations and I only include
simulations starting with a group of ellipticals. Time is measured in
arbitrary computer units}
\label{fig:correl_E}
\end{figure}

\begin{figure}
  \includegraphics[height=.3\textheight]{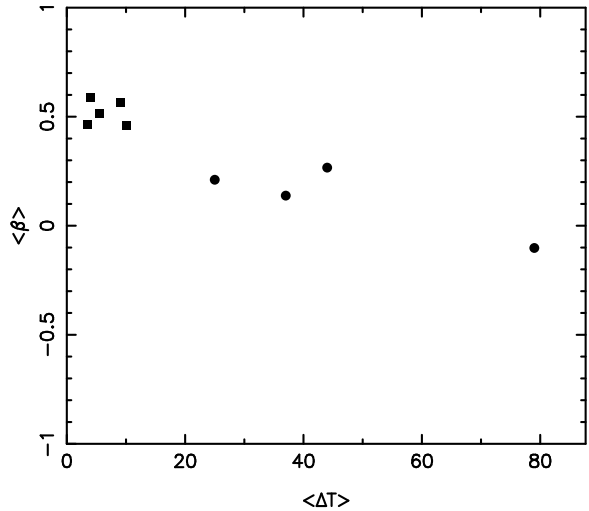}
  \caption{As in Fig.~\ref{fig:correl_E}, but for simulations starting
    with a group of disc galaxies. }
\label{fig:correl_D}
\end{figure}

\begin{figure}
  \includegraphics[height=.3\textheight]{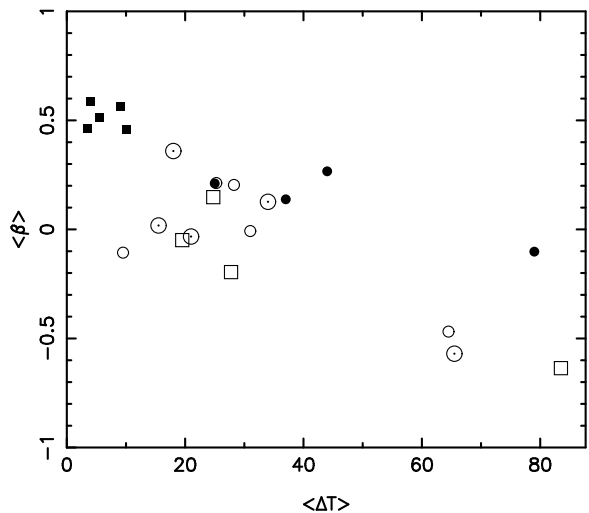}
  \caption{As in Fig.~\ref{fig:correl_E}, but for all
  simulations. Open symbols denote simulations with elliptical
  progenitors and filled ones simulations with disc galaxy progenitors.} 
\label{fig:correl_all}
\end{figure}

So what causes these differences in the orbital structure of the
remnant? Since I have a number of 
simulations, I decided to search for the parameter(s) in the initial
conditions, or in the evolutionary history that would best account for
the resulting orbital distribution. After a number of trials, I found
that the average time between consecutive mergings was the best
indicator. This is part of the evolutionary history of the
group. Indeed, if this time is large, then the merging history can be
considered as a sequence of pair mergings, initially D + D, and then D
+ E, or E + E. Inversely, if the average time between mergings is
short, then the next merging occurs before the previous one has had
time to settle and sometimes even when it is still on-going. 

In Figures~\ref{fig:correl_E}, \ref{fig:correl_D} and
\ref{fig:correl_all}, I plot $<\beta>$ as a function of the average
time between mergings, for simulations starting with ellipticals,
simulations starting with discs and all simulations together,
respectively. The first two plots show a very clear trend. 
In simulations in which the mergings followed each other closely
in time (i.e. have small $<\Delta T>$) $<\beta>$ is sometimes positive
(6 cases out of 11) and sometimes 
negative (5 cases out of 11), with many cases being near zero. Such
remnants should have near-isotropic velocity distributions. On the
other hand, simulations in which the mergings are widely spaced in
time (i.e. have large values of $<\Delta T>$) clearly have small
values of $<\beta>$, i.e. should have a fair fraction of near-circular
orbits.  

A similar trend between $<\beta>$ and average time between mergings is
seen in Figure~\ref{fig:correl_D}, this time for mergers within
groups of disc galaxies. Figure~\ref{fig:correl_all} includes all
simulations and shows that, although both types of merger remnants
follow a similar trend, the remnants from groups containing initially disc
galaxies always have an orbital structure more biased towards
near-radial orbits, as could have been expected from
Figure~\ref{fig:histo}.

The main purpose of the above discussion is to show that a variety of
effects can influence the orbital structure of the merger remnant,
such as the type of progenitors and the merging history. I did not
check whether any of my merger remnants presented a reasonable fit to the galaxies observed by Romanowsky \tal (2003) and by Mendez
\tal (2001), nor am I suggesting that this is the way in which
ellipticals were actually formed. Presumably both pair and multiple
mergers took place, in fractions which we can not yet determine. Also
mergers should have occurred both between equal and between unequal
mass galaxies. 

To summarise, the above discussion is just one more piece of the
puzzle which, together with all the rest, needs to be taken into
account and could contribute to elucidating this crucial and most
interesting problem.


\begin{theacknowledgments}
I thank Grazyna Stasi\'nska and Ryszard Szczerba for inviting me to
this very interesting and lively meeting and Albert Bosma, Martin
Bureau and Gary Mamon for stimulating discussions. 
\end{theacknowledgments}



\bibliographystyle{aipproc}   


\IfFileExists{\jobname.bbl}{}
 {\typeout{}
  \typeout{******************************************}
  \typeout{** Please run "bibtex \jobname" to obtain}
  \typeout{** the bibliography and then re-run LaTeX}
  \typeout{** twice to fix the references!}
  \typeout{******************************************}
  \typeout{}
 }


\end{document}